\documentclass[twocolumn,aps]{revtex4}
\usepackage{bm}
\usepackage[colorlinks=true,linkcolor=blue]{hyperref}
\usepackage{amsmath}
\usepackage{amssymb}
\usepackage{amsthm}
\usepackage{amsfonts}
\usepackage{times}
\usepackage{latexsym}
\usepackage{graphicx}
\begin{document}

\title{Evolution of Electronic Structure Across the Rare-Earth RNiO$_3$ Series}

\author{John W. Freeland}
\address{Advanced Photon Source, Argonne National Laboratory, Argonne, Illinois 60439, USA}

\author{Michel van Veenendaal}
\affiliation{Department of Physics, Northern Illinois University, DeKalb, Illinois 60115, USA}
\affiliation{Advanced Photon Source, Argonne National Laboratory, Argonne, Illinois 60439, USA}

\author{Jak Chakhalian}
\affiliation{Department of Physics, University of Arkansas, Fayetteville, Arkansas 70701, USA}

\begin{abstract}
The perovksite rare-earth nickelates, RNiO$_3$ (R=La ... Lu), are a class of materials displaying a rich phase-diagram of metallic and insulating phases associated with charge and magnetic order. Being in the charge transfer regime, Ni$^{3+}$ in octahedral coordination displays a strong hybridization with oxygen to form $3d$ - $2p$ mixed states, which results in a strong admixture of $3d^8\underline{L}$ into $3d^7$, where $\underline{L}$ denotes a hole on the oxygen. To understand the nature of this strongly hybridized ground state, we present a detailed study of the Ni and O electronic structure  using high-resolution soft X-ray Absorption Spectroscopy (XAS). Through a comparison of the evolution of the  XAS line-shape at Ni L- and O K-edges across the phase diagram, we explore the changes in the electronic signatures in connection with the insulating and metallic phases that support the idea of hybridization playing a fundamental role.
\end{abstract}

\date{\today}
\maketitle

\section{Introduction}
\label{intro}
Complex oxides offer a diverse range of phenomena that span magnetism to superconductivity to collosal response at phase transitions\cite{Goodenough:2004hm,Dagotto:2005ip}. Of particular interest here is understanding why materials undergo a conversion from metallic to insulating phases\cite{Imada:1998er}. Nickelates form an intriguing case for materials that undergo a metal-to-insulator transition (MIT) connected to the onset of charge and magnetic order. As is encountered in other complex oxides with multiple degrees of freedom, the underlying cause can be difficult to unravel\cite{Medarde:1997wi,Catalan:2008wb}.
\begin{figure}[h]\vspace{-0pt}
\includegraphics[width=0.45\textwidth]{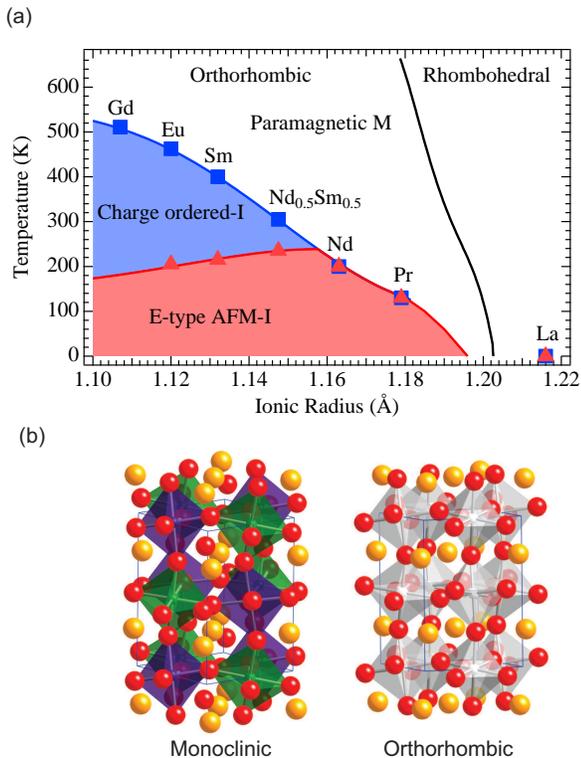} 
\caption{\label{Fig1} (a) Bulk RNiO$_3$ phase diagram for the series of samples explored in this paper. Data is from Ref. \cite{Zhou:2004jb} (b) Crystal structures for the low-temperature charge-ordered monoclinic and high-temperature metallic orthorhombic phases.}
\end{figure}

The case of RNiO$_3$ has been well studied for several decades as a case of charge and magnetic order associated with a metal to insulator transition (MIT) for  the  late 3d transition metal oxides\cite{Wold:1957gi,Goodenough:1965ww,Demazeau:1971br,Vassiliou:1989cv,Torrance:1992ir}. Originally, it was expected that the Jahn-Teller active Ni$^{3+} (3d^7)$ state should demonstrate orbital order\cite{RodriguezCarvajal:1998dy,Medarde:1998ew}, but this was ruled out in the favor of a charged ordered (CO) phase both experimentally\cite{Staub:2002cp,Scagnoli:2005je} and theoretically\cite{Mizokawa:2000wq,Mazin:2007jx,Park:2012hg,Johnston:2014tv}. The magnetic state was characterized early on as E$^{\prime}$-type antiferromagnetic (AFM) order with a 4$\times$4$\times$4 monoclinic unit cell with large and small moment Ni sites arrange in an -up-up-down-down- pattern\cite{GarciaMunoz:1992dj,RodriguezCarvajal:1998dy}; a definitive assigning the moment orientations has proven to be difficult with powder samples. Later, scattering experiments  on  relaxed NdNiO$_3$ thin films measured at the Ni L$_3$ resonance supported a spiral structure\cite{Scagnoli:2006ja,Scagnoli:2008iu} and when extended to powder samples showed the same resonant response in the soft X-rays possibly indicating the same magnetic structure\cite{Staub:2007gr,Bodenthin:2011bn}. While scattering probes showed a clear ordering of the local magnetic moments, magnetic properties explored by other methods illustrated a complex cross-over of the magnetic state from Pauli like to Curie-Weiss type\cite{Zhou:2003bf}. Recent total moment methods found the evidence for spin-canting\cite{Kumar:2013be}. Additionally, experiments on LaNiO$_3$ emphasized the difficulty in understanding the total magnetic moment  due to strong electron-electron correlations\cite{Zhou:2014er}.

On the electronic side, the trends across the series were connected with the decreasing band-width as one moves to smaller R ions\cite{Torrance:1992ir,Sarma:1994eo,Zhou:2004jb,Zhou:2004fv}, which was also early on associated with a change in the $p-d$ covalency\cite{Barman:1994ee} that has a consequence of moving the boundary between metal and insulator in the Zaanen-Sawatzky-Allen  (ZSA) diagram for charge-transfer compounds\cite{Zaanen:1985wp}. Photoemission\cite{Medarde:1997bu,Vobornik:1999de,Okazaki:2003to,Schwier:2012gqa} and optical\cite{Katsufuji:1995dy} measurements also showed that the gap is quite small (100 - 200 meV). Early on it was recognized that the charge-transfer nature of the compounds played an important role in the physics\cite{Mizokawa:1995vq,Medarde:1997wi}, which indicated that Ni$^{3+}$ contained more  of $3d^8\underline{L}$ character than $3d^7$, where $\underline{L}$ denotes a ligand hole on the oxygen site. This was explored in more detail by Mizokawa et al.\cite{Mizokawa:2000wq} and further supported by  recent advances in theoretical approaches for strongly correlated electron states\cite{Mazin:2007jx,Lee:2011bq,Park:2012hg,Johnston:2014tv,Subedi:2015en}. These findings support the  view that the charge order is more likely \textit{oxygen-site} or bond centered ($3d^8\underline{L}$ + $3d^8\underline{L}$ $\rightarrow$ $3d^8$ + $3d^8\underline{L}^2$) rather than the Ni site centered picture ($3d^7$ + $3d^7$ $\rightarrow$ $3d^{7-\delta}$ +$3d^{7+\delta}$).

In this article, we explore the detailed evolution of the nickel and oxygen electronic structure throughout the phase diagram  above and below the MIT and AFM ordering transitions by using spectroscopic probes at the Ni L-edge and Oxygen K-edge to track the evolution of the electronic structures to understand  the underlying physics.

\section{Experimental Details}
\label{expt}
Bulk RNiO$_3$ samples were studied in the powder form and the details of high-pressure oxygen synthesis are discussed elsewhere\cite{Zhou:2004fv}. X-ray absorption measurements were undertaken at beamline 4-ID-C of the Advanced Photon Source and were simultaneously recorded  by surface sensitive total electron yield (EY) and bulk-sensitive partial fluorescence yield (FY) modes. In addition, a NiO sample was positioned upstream of the experiment and recorded during every scan to perform alignment between samples with up to 50 meV precision. Powder
samples were mounted on electrically conducting carbon tape. Excellent agreement between EY and FY measurements indicates no degradation of the surface  with respect to the bulk phase. The beamline resolution was 100 meV at the oxygen K-edge and 200 meV at the Ni-L edge.
\begin{figure}[h]\vspace{-0pt}
\includegraphics[width=0.45\textwidth]{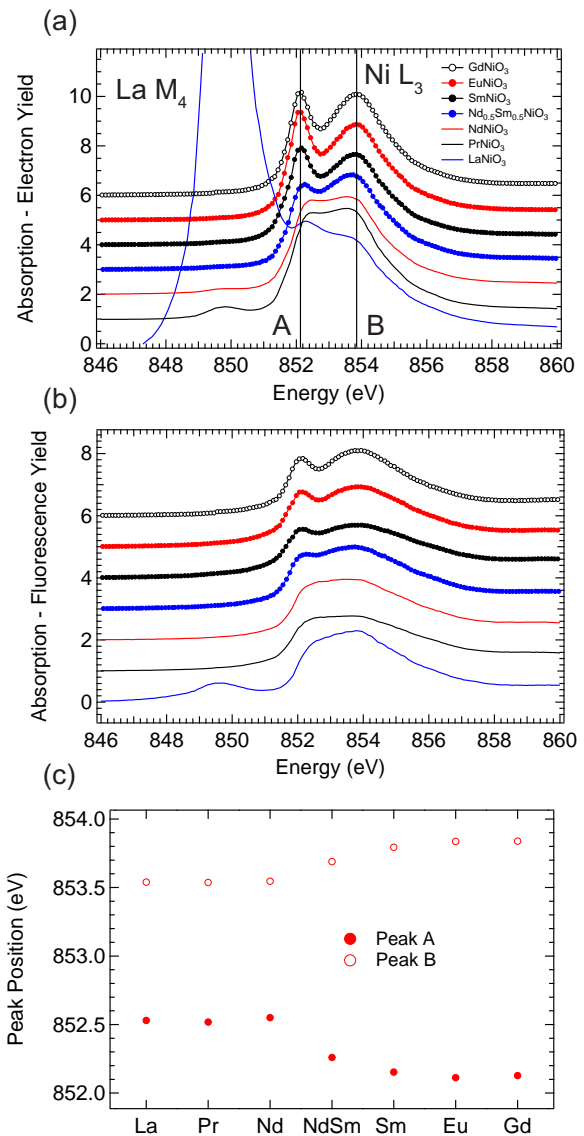} 
\caption{\label{Fig2} Ni L$_3$ absorption data at 300 K acquired in (a) surface senstive EY and (b) bulk sensitive FY across the series from R = Gd to La. Panel (c) shows the multiplet peak splitting that is discussed in the text.}
\end{figure}

\section{Results}
\subsection{Ni L-edge Absorption}
\label{nil}
In this section we present a summary of the Ni L-edge X-ray absorption, which is an excitation from the Ni {\it 2p} core state to the unoccupied Ni 3$d$ states just above the Fermi level. Figure \ref{Fig2} shows the L$_3$ portion of the spectra in both EY (a) and FY (b). Aside from the strong self-absorption of the FY spectra, there is a very good correspondence of the spectral features across the series from R = Gd to La. Aside from the overlap with the strong La M$_4$ line for LaNiO$_3$, the spectra show a clear evolution across the series as one moves across the metal-insulator transition. These spectra are in good agreement with previous measurements on bulk RNiO$_3$\cite{Mizokawa:1995vq,Piamonteze:2005ux}. The spectra are qualitatively described by a sharp mutliplet near 852 eV and a broad multiplet near 854 eV, labeled as A and B respectively in Fig.\ \ref{Fig2}. The bulk sensitive FY shows the same features indicating that the surface and bulk have the same electronic structure although they are less pronounced due to strong self-absorption at the L$_3$ edge\cite{vanderLaan:2014if}. Note that these spectra were aligned using a NiO standard which has a peak at 852 eV, which is close yet distinct from the multiplet labeled A. While not presented here, a change in valence from Ni$^{3+}$ to Ni$^{2+}$ due to oxygen non-stoichiometry (e.\ g.\ RNiO$_{3-\delta}$) occurs by an increase in the A peak intensity and gradual shifting to 852 eV while the B peak diminshes.

By treating the spectra as containing two main components, we have fit all the L$_3$ with a two Gaussian model to extract the position of the two main multiplets. While the multiplets are clear in the insulating phase, there also exists features in the metallic state  consistent with some remaining multiplet character, which is heavily screened by the metallic state. The results of this fit can be seen in Fig.\ \ref{Fig2}(c) as a function of rare earth ion. A direct inspection of the Fig. shows the presense of  two distinct well defined splittings depending on the metallic vs.\  insulating nature of the ground state of 1 eV (metallic) and 1.7 eV (insulating), respectively. The data of Nd$_{0.5}$Sm$_{0.5}$ is an outlier since at 300 K and are very close to the transition from the metallic to insulating phase. In previous work, we  used a NiO$_6$ cluster calculation that connected the splitting of these features to the charge transfer energy, $\Delta^{(3)}\ =\ E(3d^8\underline{L})\ -\ E(3d^7)$\cite{Liu:2011ej}. This notion will be addressed in more detail in the Theory and Discussion sections below, and  implies that there might be a discontinuous change in $\Delta$ across the MIT.
\begin{figure}[h]\vspace{-0pt}
\includegraphics[width=0.5\textwidth]{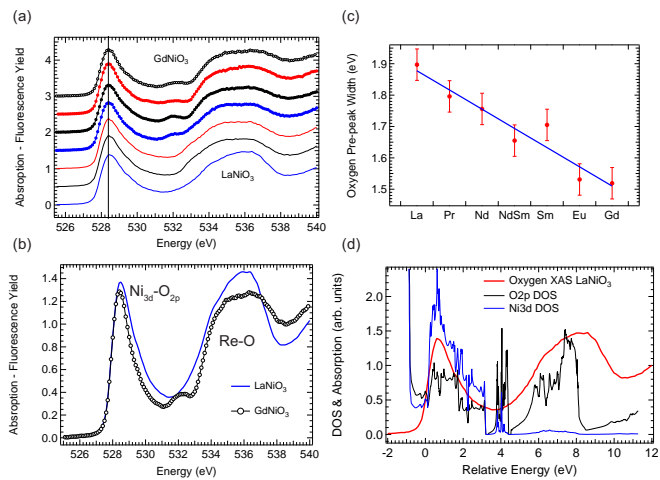} 
\caption{\label{Fig3} (a) The oxygen K-edge spectra at 300K for the series from R = Gd to La with the Ni$_{3d}$-O$_{2p}$ marked by the vertical. (b) shows a comparison of the ends members of the measured series highlighting the pre-peak from hybridization with Ni and the second feature due Re-O hybridization. (c) This panel shows the FWHM of the pre-peak at 300K as a function of R. (c) Comparison between LaNiO$_3$ XAS and the oxygen projected density of states (DOS) from Ref. \cite{Gou:2011ky}}
\end{figure}

\subsection{Oxygen K-edge Absorption}
\label{ok}
Due to the highly covalent nature of the bonding between Ni and O, probing the oxygen states is equally important to that of Ni. Figure\ \ref{Fig3} shows the oxygen K-edge absorption spectrum for the series shown in Fig.\ \ref{Fig1}(a). In all cases the spectra are broadly defined by a sharp pre-peak around 529 eV due to Ni $3d$ and O $2p$ hybridization and a broad peak around 538 eV connected to mixing of the rare earth (R) states with oxygen. These spectra are in good agreement with other published data\cite{Medarde:1992ds,Mizokawa:1995vq,Abbate:2002ti,Suntivich:2014ft} and are consistent with fully oxygenated case of RNiO$_3$\cite{Abbate:2002ti}. In this case we present only the bulk-sensitive FY since they suffer less from self-absorption and due the fact that the EY data for the insulating samples lead to difficulties in creating a consistent comparison between the data.

One of the main findings of this measurement is that the leading edge of the pre-peaks are all aligned and the changes occur only the higher energy side of the pre-peak. This is best seen in the comparison of the two ends of the series for Gd and La shown in Figure\ \ref{Fig3}(b). Here we see that the main effect is the reduction in the pre-peak width and perhaps a drop in overall area size. Since small changes in the normalization of the oxygen spectra can influence the small changes in the pre-peak height, here we focus on the changes in the peak width across the series (Fig.\ \ref{Fig3}(c)). As clearly seen, there is a linear trend that tracks well with the decreasing bandwidth across the RNiO$_3$ series inferred from changes in the bond angle\cite{Torrance:1992ir,Zhou:2004jb}. With the shrinking $3d$ bandwidth, the overlap with the $2p$ bands decreases as seen by the loss of pre-peak intensity at higher energy.  An important aspect of the oxygen K-edge measurements is that due to the weak core-hole interaction at this edge, the XAS can be directly compared to the oxygen partial density of states\cite{Sarma:1996bs,deGroot:2005wg,Medling:2012wm}. To illustrate this point we consider the comparison of the LaNiO$_3$ XAS with the calculated density of states from Ref.\ \cite{Gou:2011ky}. As seen, there is  the good agreement between the two  that will be utilized during the discussion in connection to the underlying physics.

\subsection{Case of NdNiO$_3$}
\label{nno}
\begin{figure}[h]\vspace{-0pt}
\includegraphics[width=0.45\textwidth]{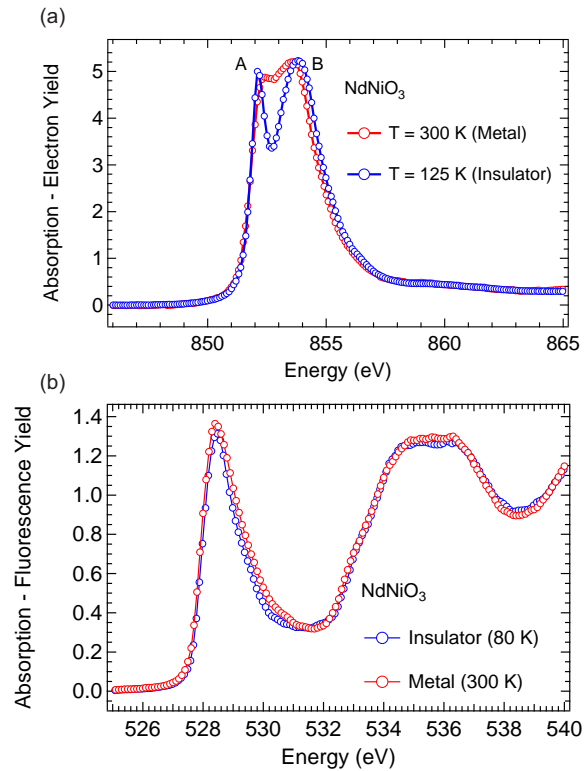} 
\caption{\label{Fig4} (a) Ni L$_3$ edge below and above the MIT. (b) Corresponding oxygen K-edge in the metallic vs.\ insulating phases.}
\end{figure}
The situation of R = Nd offers a representative compound to explore  the spectroscopy of the insulating vs.\  metallic phases. As shown in Fig.\ \ref{Fig4}, as a function of temperature the Ni L-edge across the MIT changes in the same manner  as it does as a function of R. Specifically, below the MIT (T$_{MIT}$=190 K) there is an abrupt change in the XAS into the multiplet split phase of the Ni L-edge for NdNiO$_3$, which is exactly the same as that of the other insulating phases. In this context, we can conjecture that understanding the spectra for the case of Nd should provide an understanding of all the other members in the series. For the case of oxygen, tracking the change in the same compound allows a quantitative comparison of the variation in the oxygen spectrum. As seen in Fig. 4, there is a narrowing of the pre-peak associated with the decrease of the Ni $3d$ bandwidth that results in a decrease of the overlap with the oxygen $2p$ states. The integrated area decreases by $\sim$ 15\% in the insulating phase. A similar  analysis of the Ni L-edge is not as clear largely due to the change of the screening across the transition that complicates analysis of the total area of the spectrum.

At this point is worth noting that the same change occurs for the case of PrNiO$_3$ albeit at a lower temperature while for LaNiO$_3$ no noticeable change found in the L-edge spectrum at low temperature, which is consistent with the persistent metallic phase. For the case of R from Sm to Gd, we also looked for changes when crossing the low temperature transition into the E'-type AFM phase; no clear changes has been observed at the  L-edge. This indicates that the fundamental change in the L-edge spectrum is connected with the transition into the insulating phase rather than to magnetism. In addition, the oxygen K-edge shows a very slight decrease in the pre-peak that will be discussed in more detail below.

\subsection{Theory of the Ni L-edge X-ray Absorption Lineshape}
\label{theory}
The ground state of trivalent nickel compounds has been a long-standing discussion \cite{Fujimori:1984ju,Kuiper:1989ep,VanElp:1992ko,VanVeenendaal:1993dx,Mizokawa:1995vq,Piamonteze:2005ux}. The question is directly related to the lowest electron-removal state of divalent nickel compounds. Historically Ni(2+)O has been extensively  studied in context of charge transfer insulators. The ground state of NiO is a high-spin triplet state with two holes in the $e_g$ orbitals. Within octahedral symmetry ($O_h$), the ground state is $^3A_2$ and can be written as $|\underline{d}_{x^2-y^2\uparrow}\underline{d}_{3z^2-r^2\uparrow}\rangle$. NiO is known as a charge-transfer insulator with a gap of 4.3 eV\cite{Sawatzky:1984jt}. The size of the insulating gap is directly related to the charge-transfer energy $\Delta^{(n)}$ for $n$ holes, defined as
\begin{equation}
\Delta^{(2)}=E(d^9{\underline L} )-E(d^8),
\end{equation}
where  ${\underline L}$ stands for a hole on the oxygen ligands. In order to reproduce the gap, a charge-transfer energy of 5.5-6.5 eV is needed. Since the Coulomb repulsion is of the same size, NiO is in fact very close to the boundary between a charge-transfer insulator and a Mott-Hubbard insulator \cite{Zaanen:1985wp}. This observation has important consequences for trivalent nickel compounds.  Instead of a high-spin state favored  by the Hund's exchange terms in the $dd$ Coulomb interaction \cite{Fujimori:1984ju}, the strong covalency stabilizes  the low-spin state ($^2E$ in $O_h$) \cite{VanElp:1992ko}.   The strong covalency is due to the lowering of the  charge-transfer energy when the valency is increased. The charge-transfer energy for the trivalent system can, in a first approximation, be related to that of the divalent system via
 \begin{equation}
\Delta^{(3)}=E(d^8{\underline L} )-E(d^7)\cong \Delta^{(2)}-U.
\end{equation}
 This situation is  similar to copper oxides, which, with a charge-transfer energy of 2-3 eV, are clearly in the charge-transfer regime. For cuprates, the lowest electron-removal state is a singlet \cite{Zhang:1988jf,Eskes:1988ef} with predominantly $d^9{\underline L}$ character. When hole is doped in to a cuprate, these lowest electron-removal states, known as a Zhang-Rice singlet, are responsible for the conductivity. For nickelates the configuration of the lowest electron-removal state has mainly $|\underline{d}^2_{e_g\uparrow}\underline{L}_{e_{g}\downarrow} \rangle$ character. The assignment of low-spin was further supported by O $1s$ X-ray absorption \cite{Kuiper:1989ep} that clearly showed a strong increase in the ligand-hole character with hole doping of NiO via Li substitution. In addition, the $2p\rightarrow 3d$ X-ray absorption could only be interpreted following the assumption that the additional holes couple antiparallel to the spin on the nickel ion \cite{VanVeenendaal:1994kh}. 
\begin{figure}[h]\vspace{-0pt}
\includegraphics[width=0.45\textwidth]{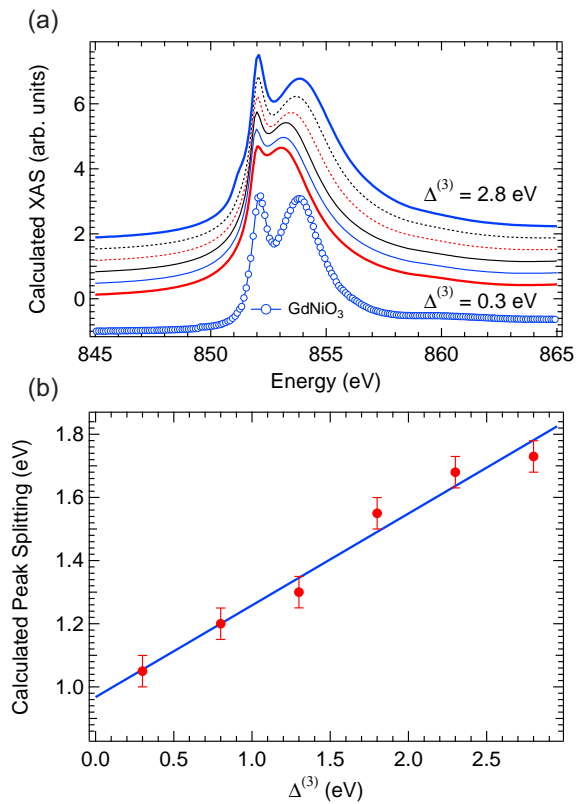} 
\caption{\label{Fig5} (a) Calculated absorption spectra for Ni$^{3+}$ in a NiO$_6$ cluster as a function of the charge transfer energy, $\Delta^{(3)}$. Included is a comparison with the Ni L-edge of GdNiO$_3$ in the insulating phase. (b) The A-B peak splitting described in Fig.\ \ref{Fig2} extracted from the calculated spectra as a function of $\Delta^{(3)}$. The  line is a linear fit to the results. }
\end{figure}

%The competition between low-spin and high-spin nickel compounds is also relevant to trivalent nickel compounds, such as $R$NiO$_3$. It is important the charge-transfer energy for these systems is not necessarily equal to $\Delta^{(3)}$ due to changes in the chemical environment. 

Following this idea, calculations were performed for a NiO$_6$ cluster with octahedral coordination using the methods described in Ref. \cite{VanVeenendaal:1993dx}. The Hamiltonian includes the on-site Coulomb between the $3d$ electrons and between the $3d$ electrons and the $2p$ core hole. The parameters are calculated within the Hartree-Fock limit and scaled down to 80\% to account for intra-atomic screening effects. The monopole parts were $F^0_{dd}=6$ eV and $F^0_{pd}=7$ eV. The spin-orbit coupling was included for the $3d$ and $2p$ electrons. The hybridization with the ligands was taken into account by including configurations up to double ligand hole. The hybridization parameter is $V=2.25, -1.03$ eV for the $e_g$ and $t_{2g}$  orbitals, respectively. The cubic crystal field $10Dq$ is 1.5 eV.

Despite its relative simple appearance, the $L$-edge spectra for trivalent nickel compounds are significantly more difficult to interpret that those for divalent nickel systems, such as NiO. As seen in Fig ???, the spectra consist of a sharp peak at the low-energy side of the edge, followed at higher energy by a broad feature. These broadenings are difficult to explain by multiplet effects with a constant broadening and indicate the presence of additional broadening effects. First, it is important to note that the sharp features often found at the $L$-edges in  transition-metal compounds are due to the creation of an excitonic state below the continuum states by the strong Coulomb attraction between the $2p$ core hole and the  $3d$ electrons. 
The energy needed to transfer an electron from the ligands to the transition-metal in the XAS final states is given $\Delta^{(n)}-U_c+U$. Since generally $U_c>U$, the effective charge-transfer is lower in the XAS final states compared to the ground state. For divalent nickel oxides, such as NiO, $\Delta^{(2)}\cong 5.5-6.5$ eV, and this difference is not too important. However, for trivalent nickel oxides the charge-transfer energy is approximately $\Delta^{(3)}\cong 0-3$ eV. This means that, in particular, the high-energy final states can overlap with the continuum states.  The dramatic effects of coupling to the continuum were apparent in non-resonant inelastic x-ray scattering on rare-earth compounds \cite{Gordon:2008hb}, where  the substantial broadening     in addition to the usual core-hole lifetime broadening was observed for states at higher energies. This  coupling to the continuum can be included by having the Lorentzian broadening increase parabolically from the core-hole broadening due to Auger effects of 0.25 eV at the absorption edge to 1.25 eV at 1 eV above the edge. Assuming that the coupling to the continuum does not vary strongly this broadening is then kept constant. 

Figure \ref{Fig5} shows the change in spectral line shape as a function of charge-transfer energy. Decreasing the charge-transfer energy makes the system more covalent. In the final state, the $\underline{1s}3d^9\underline{L}$ states will be closer in energy to the $\underline{1s}3d^8$ configurations. The closer proximity increases the hybridization between these states leading to a compression of the spectral features\cite{VanDerLaan:1986ez}. At the $L_3$ edge, the decrease in covalency increases the energy separation between the two features. At the $L_2$ edge, an increase in charge-transfer energy leads to the development of a more pronounced shoulder at the low-energy side. These trends correspond well to the trends observed experimentally in Ref. \cite{Piamonteze:2005ux} and in Fig. \ref{Fig5}. In metallic systems with $RE=$Nd or Pr at low temperatures, the covalency is stronger and no clear separation of the features in the $L_3$ edge is observed. For insulators with $R=$Eu, Y, Lu, on the other hand, the $L_3$ is clearly split into two features.

\section{Discussion}
\label{diss}
In this section, we begin with a quick summary of the key features observed in our measurements. As shown above, one of the key question concerns understanding the role of the $3d^8\underline{L}$ character in the ground state of nickelates and the data analysis requires to  include this state to  reproduce the observed XAS\ features. As noted by Mizokawa et al.\cite{Mizokawa:2000wq}, nickelates should really be view as a self-doped Mott insulator, where the doping comes from the oxygen $2p$ band\cite{Ushakov:2011fy}. Certainly understanding the spectroscopy necessitates the inclusion of a significant $3d^8\underline{L}$ component to reproduce a similar spectrum from the calculations. As seen above in the Ni L-edge data, the key factor in the change of the spectrum shown in Figs.\ \ref{Fig2} and \ref{Fig4} is the transition from the metallic to insulating phase. In the calculations shown in Figure \ref{Fig5}, the splitting between the A and B multiplets is directly influenced by the charge transfer energy, $\Delta^{(3)}$. An interesting suggestion is that the change between the features in the metallic vs.\  insulating phases could be due to a change in $\Delta^{(3)}$. Since the charge transfer energy is closely tied to the splitting of the $3d$ and $2p$ levels, perhaps as a gap opens there is an increase in the offeset of these states. However, the small nature of the gap\cite{Katsufuji:1995dy,Medarde:1997bu,Vobornik:1999de,Okazaki:2003to,Schwier:2012gqa} might suggest that screening in the metallic phase also plays a role since there is a Madelung term in the derivation of the charge transfer energy\cite{Ohta:1991ft} that could be affected by free-carriers and covalency.

\begin{figure}[h]\vspace{-0pt}
\includegraphics[width=0.45\textwidth]{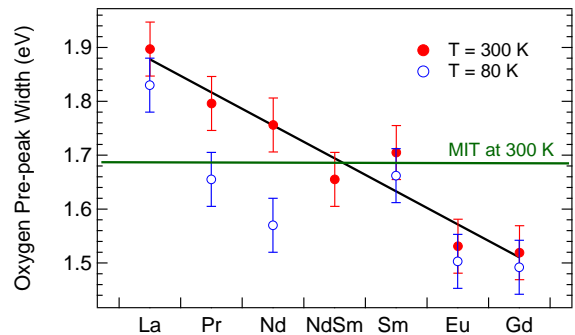} 
\caption{\label{Fig6} The Ni$_{3d}$-O$_{2p}$ oxygen pre-peak width as a function of R and temperature. The green line marks the extrapolated width for the case of an MIT at 300K.}
\end{figure}
From the oxygen site perspective, the strong pre-peak also indicates a significant $3d^8\underline{L}$ contribution to the ground-state wave function. Unlike the Ni L-edge data that undergoes a profound change in lineshape, the K-edge is much more subtle and in line with the energy scale of the physics that splits the metallic vs. insulating phases. Within 50 meV, there is no observable increase in the energy of the leading edge that has been observed in other systems where gaps open\cite{Cavalleri:2004kf,Cavalleri:2005fe,Merz:2006jd,Rini:2009cg}. For the case of oxygen all of the changes occur on the high energy side and result in a decreased peak-width (and correspondingly peak area) consisted with the reduction in $d-p$ overlap resulting from a bandwidth reduction. Since the peak area is proportional to the $3d^8\underline{L}$ component, this indicates the reduction in the ligand hole content both as a function of R and across the MIT. This is clearly seen in Fig.\ \ref{Fig6}, where we have include the low temperature data for the pre-peak width. For the cases where the system always stay metallic, there is also a slight decrease in the width, although it is just outside the error bars  and might not be real. For the R = Sm, Eu, Gd, there is no change when crossing the magnetic transition indicating the primary effect in the bandwidth reduction occurs at the MIT. For the cases of Pr and Nd, there is clear drop in the pre-peak width associated with a reduction of $3d^8\underline{L}$ spectral weight.

While band effects play a very important role here,  by starting from the ionic perspective the wave function can be written as:
 \begin{equation}
 |Ni^{3+}\rangle\ =\ \alpha|3d^7\rangle\ +\ \beta|3d^8\underline{L}\rangle\ +\ \gamma|3d^9\underline{L}^2\rangle\ +\ \dots
\end{equation}
which implies that a reduction in the ligand hole terms will lead to an increase in the $3d^7$ component. This is basically the idea first promoted by Sarma et al.\ \cite{Barman:1994ee}, that changes in the $p-d$ hybridization can move the metal to insulator line in the ZSA diagram. More recent work with DFT-DMFT\cite{Wang:2012dd}, presented a similar diagram from the perspective of the local count of the $3d$ electrons, $N_d$. In this self-doped case, a reduction in the ligand hole components could reduce $N_d$ and push the system across the MIT boundary if it were at the close proximity. From the perspective of this picture, one might imagine that there is a critical value of the ligand hole weight below which the system becomes insulating as highlighted by the green line in Fig.\ \ref{Fig6}. This also connects with the recent results that suggest there could be a \textit{pre-formed} bond-disproportionated state even in the metallic phase\cite{Piamonteze:2005cd,Medarde:2009hg,Jaramillo:2014db}.  In the future, we hope to develop better ways to model these spectra in order to  quantitatively assess these changes to better understand the physics of the MIT in the rare-earth nickelates and other complex oxides in the charge-transfer regime.

\section{Conclusion}
\label{conc}
In conclusion, we have highlighted the importantance if the $3d^8\underline{L}$  state  in order to understand the details of the Ni L- and O K- edge spectra of the perovskite nickelates. Given the detailed knowledge that exists about the physical properties as a function of rare earth ion, R, we hope that the systematics of this data can aid in building a better picture of the physics of co-existing charge and magnetic orders that result in a metal to insulator transition.

\section{Acknowledgements}
\label{ack}
This article is dedicated to D.D. Sarma whose discusssions, insight and invaluable comments concerning this data over the last several years have made this paper possible. The authors thank Bogdan Dabrowski for providing the high-quality RNiO$_3$ samples. Work at the Advanced Photon Source, Argonne National Laboratory was supported by the U.S. Department of Energy, Office of Science under Grant No. DEAC02-06CH11357. MvV was supported by the U. S. Department of Energy (DOE), Office of Basic Energy Sciences, Division of Materials Sciences and Engineering under Award No. DE-FG02-03ER46097 and NIU's Institute for Nanoscience, Engineering, and Technology.
The work at the University of Arkansas is funded in part by the Gordon and Betty Moore Foundation's EPiQS Initiative through Grant GBMF4534 and by the DOD-ARO under Grant No. 0402-17291. 


%% References

\bibliography{BulkRNO_bib.bib}

\begin{thebibliography}{69}
\expandafter\ifx\csname natexlab\endcsname\relax\def\natexlab#1{#1}\fi
\expandafter\ifx\csname bibnamefont\endcsname\relax
  \def\bibnamefont#1{#1}\fi
\expandafter\ifx\csname bibfnamefont\endcsname\relax
  \def\bibfnamefont#1{#1}\fi
\expandafter\ifx\csname citenamefont\endcsname\relax
  \def\citenamefont#1{#1}\fi
\expandafter\ifx\csname url\endcsname\relax
  \def\url#1{\texttt{#1}}\fi
\expandafter\ifx\csname urlprefix\endcsname\relax\def\urlprefix{URL }\fi
\providecommand{\bibinfo}[2]{#2}
\providecommand{\eprint}[2][]{\url{#2}}

\bibitem[{\citenamefont{Goodenough}(2004)}]{Goodenough:2004hm}
\bibinfo{author}{\bibfnamefont{J.}~\bibnamefont{Goodenough}},
  \bibinfo{journal}{Reports On Progress In Physics}
  \textbf{\bibinfo{volume}{67}}, \bibinfo{pages}{1915} (\bibinfo{year}{2004}).

\bibitem[{\citenamefont{Dagotto}(2005)}]{Dagotto:2005ip}
\bibinfo{author}{\bibfnamefont{E.}~\bibnamefont{Dagotto}},
  \bibinfo{journal}{Science} \textbf{\bibinfo{volume}{309}},
  \bibinfo{pages}{257} (\bibinfo{year}{2005}).

\bibitem[{\citenamefont{Imada et~al.}(1998)\citenamefont{Imada, Fujimori, and
  Tokura}}]{Imada:1998er}
\bibinfo{author}{\bibfnamefont{M.}~\bibnamefont{Imada}},
  \bibinfo{author}{\bibfnamefont{A.}~\bibnamefont{Fujimori}}, \bibnamefont{and}
  \bibinfo{author}{\bibfnamefont{Y.}~\bibnamefont{Tokura}},
  \bibinfo{journal}{Reviews Of Modern Physics} \textbf{\bibinfo{volume}{70}},
  \bibinfo{pages}{1039} (\bibinfo{year}{1998}).

\bibitem[{\citenamefont{Medarde}(1997)}]{Medarde:1997wi}
\bibinfo{author}{\bibfnamefont{M.}~\bibnamefont{Medarde}},
  \bibinfo{journal}{JOURNAL OF PHYSICS CONDENSED MATTER}
  (\bibinfo{year}{1997}).

\bibitem[{\citenamefont{Catalan}(2008)}]{Catalan:2008wb}
\bibinfo{author}{\bibfnamefont{G.}~\bibnamefont{Catalan}},
  \bibinfo{journal}{Phase Transitions} \textbf{\bibinfo{volume}{81}},
  \bibinfo{pages}{729} (\bibinfo{year}{2008}).

\bibitem[{\citenamefont{Zhou and Goodenough}(2004)}]{Zhou:2004jb}
\bibinfo{author}{\bibfnamefont{J.}~\bibnamefont{Zhou}} \bibnamefont{and}
  \bibinfo{author}{\bibfnamefont{J.}~\bibnamefont{Goodenough}},
  \bibinfo{journal}{Physical Review B} \textbf{\bibinfo{volume}{69}},
  \bibinfo{pages}{153105} (\bibinfo{year}{2004}).

\bibitem[{\citenamefont{Wold et~al.}(1957)\citenamefont{Wold, Post, and
  Banks}}]{Wold:1957gi}
\bibinfo{author}{\bibfnamefont{A.}~\bibnamefont{Wold}},
  \bibinfo{author}{\bibfnamefont{B.}~\bibnamefont{Post}}, \bibnamefont{and}
  \bibinfo{author}{\bibfnamefont{E.}~\bibnamefont{Banks}},
  \bibinfo{journal}{Journal Of The American Chemical Society}
  \textbf{\bibinfo{volume}{79}}, \bibinfo{pages}{4911} (\bibinfo{year}{1957}).

\bibitem[{\citenamefont{Goodenough and Raccah}(1965)}]{Goodenough:1965ww}
\bibinfo{author}{\bibfnamefont{J.~B.} \bibnamefont{Goodenough}}
  \bibnamefont{and} \bibinfo{author}{\bibfnamefont{P.~M.}
  \bibnamefont{Raccah}}, \bibinfo{journal}{Journal Of Applied Physics}
  \textbf{\bibinfo{volume}{36}}, \bibinfo{pages}{1031} (\bibinfo{year}{1965}).

\bibitem[{\citenamefont{Demazeau et~al.}(1971)\citenamefont{Demazeau, Marbeuf,
  Pouchard, and Hagenmuller}}]{Demazeau:1971br}
\bibinfo{author}{\bibfnamefont{G.}~\bibnamefont{Demazeau}},
  \bibinfo{author}{\bibfnamefont{A.}~\bibnamefont{Marbeuf}},
  \bibinfo{author}{\bibfnamefont{M.}~\bibnamefont{Pouchard}}, \bibnamefont{and}
  \bibinfo{author}{\bibfnamefont{P.}~\bibnamefont{Hagenmuller}},
  \bibinfo{journal}{Journal of Solid State Chemistry}
  \textbf{\bibinfo{volume}{3}}, \bibinfo{pages}{582} (\bibinfo{year}{1971}).

\bibitem[{\citenamefont{Vassiliou et~al.}(1989)\citenamefont{Vassiliou,
  Hornbostel, Ziebarth, and Disalvo}}]{Vassiliou:1989cv}
\bibinfo{author}{\bibfnamefont{J.~K.} \bibnamefont{Vassiliou}},
  \bibinfo{author}{\bibfnamefont{M.}~\bibnamefont{Hornbostel}},
  \bibinfo{author}{\bibfnamefont{R.}~\bibnamefont{Ziebarth}}, \bibnamefont{and}
  \bibinfo{author}{\bibfnamefont{F.~J.} \bibnamefont{Disalvo}},
  \bibinfo{journal}{Journal of Solid State Chemistry}
  \textbf{\bibinfo{volume}{81}}, \bibinfo{pages}{208} (\bibinfo{year}{1989}).

\bibitem[{\citenamefont{Torrance et~al.}(1992)\citenamefont{Torrance, Lacorre,
  Nazzal, Ansaldo, and Niedermayer}}]{Torrance:1992ir}
\bibinfo{author}{\bibfnamefont{J.}~\bibnamefont{Torrance}},
  \bibinfo{author}{\bibfnamefont{P.}~\bibnamefont{Lacorre}},
  \bibinfo{author}{\bibfnamefont{A.}~\bibnamefont{Nazzal}},
  \bibinfo{author}{\bibfnamefont{E.}~\bibnamefont{Ansaldo}}, \bibnamefont{and}
  \bibinfo{author}{\bibfnamefont{C.}~\bibnamefont{Niedermayer}},
  \bibinfo{journal}{Physical Review B} \textbf{\bibinfo{volume}{45}},
  \bibinfo{pages}{8209} (\bibinfo{year}{1992}).

\bibitem[{\citenamefont{Rodr{\'\i}guez-Carvajal
  et~al.}(1998)\citenamefont{Rodr{\'\i}guez-Carvajal, Rosenkranz, Medarde,
  Lacorre, Fernandez-D{\'\i}az, Fauth, and Trounov}}]{RodriguezCarvajal:1998dy}
\bibinfo{author}{\bibfnamefont{J.}~\bibnamefont{Rodr{\'\i}guez-Carvajal}},
  \bibinfo{author}{\bibfnamefont{S.}~\bibnamefont{Rosenkranz}},
  \bibinfo{author}{\bibfnamefont{M.}~\bibnamefont{Medarde}},
  \bibinfo{author}{\bibfnamefont{P.}~\bibnamefont{Lacorre}},
  \bibinfo{author}{\bibfnamefont{M.}~\bibnamefont{Fernandez-D{\'\i}az}},
  \bibinfo{author}{\bibfnamefont{F.}~\bibnamefont{Fauth}}, \bibnamefont{and}
  \bibinfo{author}{\bibfnamefont{V.}~\bibnamefont{Trounov}},
  \bibinfo{journal}{Physical Review B} \textbf{\bibinfo{volume}{57}},
  \bibinfo{pages}{456} (\bibinfo{year}{1998}).

\bibitem[{\citenamefont{Medarde et~al.}(1998)\citenamefont{Medarde, Lacorre,
  Conder, Fauth, and Furrer}}]{Medarde:1998ew}
\bibinfo{author}{\bibfnamefont{M.}~\bibnamefont{Medarde}},
  \bibinfo{author}{\bibfnamefont{P.}~\bibnamefont{Lacorre}},
  \bibinfo{author}{\bibfnamefont{K.}~\bibnamefont{Conder}},
  \bibinfo{author}{\bibfnamefont{F.}~\bibnamefont{Fauth}}, \bibnamefont{and}
  \bibinfo{author}{\bibfnamefont{A.}~\bibnamefont{Furrer}},
  \bibinfo{journal}{Physical Review Letters} \textbf{\bibinfo{volume}{80}},
  \bibinfo{pages}{2397} (\bibinfo{year}{1998}).

\bibitem[{\citenamefont{Staub et~al.}(2002)\citenamefont{Staub, Meijer, Fauth,
  Allenspach, Bednorz, Karpinski, Kazakov, Paolasini, and
  d'Acapito}}]{Staub:2002cp}
\bibinfo{author}{\bibfnamefont{U.}~\bibnamefont{Staub}},
  \bibinfo{author}{\bibfnamefont{G.}~\bibnamefont{Meijer}},
  \bibinfo{author}{\bibfnamefont{F.}~\bibnamefont{Fauth}},
  \bibinfo{author}{\bibfnamefont{R.}~\bibnamefont{Allenspach}},
  \bibinfo{author}{\bibfnamefont{J.}~\bibnamefont{Bednorz}},
  \bibinfo{author}{\bibfnamefont{J.}~\bibnamefont{Karpinski}},
  \bibinfo{author}{\bibfnamefont{S.}~\bibnamefont{Kazakov}},
  \bibinfo{author}{\bibfnamefont{L.}~\bibnamefont{Paolasini}},
  \bibnamefont{and}
  \bibinfo{author}{\bibfnamefont{F.}~\bibnamefont{d'Acapito}},
  \bibinfo{journal}{Physical Review Letters} \textbf{\bibinfo{volume}{88}},
  \bibinfo{pages}{126402} (\bibinfo{year}{2002}).

\bibitem[{\citenamefont{Scagnoli et~al.}(2005)\citenamefont{Scagnoli, Staub,
  Janousch, Mulders, Shi, Meijer, Rosenkranz, Wilkins, Paolasini, Karpinski
  et~al.}}]{Scagnoli:2005je}
\bibinfo{author}{\bibfnamefont{V.}~\bibnamefont{Scagnoli}},
  \bibinfo{author}{\bibfnamefont{U.}~\bibnamefont{Staub}},
  \bibinfo{author}{\bibfnamefont{M.}~\bibnamefont{Janousch}},
  \bibinfo{author}{\bibfnamefont{A.}~\bibnamefont{Mulders}},
  \bibinfo{author}{\bibfnamefont{M.}~\bibnamefont{Shi}},
  \bibinfo{author}{\bibfnamefont{G.}~\bibnamefont{Meijer}},
  \bibinfo{author}{\bibfnamefont{S.}~\bibnamefont{Rosenkranz}},
  \bibinfo{author}{\bibfnamefont{S.}~\bibnamefont{Wilkins}},
  \bibinfo{author}{\bibfnamefont{L.}~\bibnamefont{Paolasini}},
  \bibinfo{author}{\bibfnamefont{J.}~\bibnamefont{Karpinski}},
  \bibnamefont{et~al.}, \bibinfo{journal}{Physical Review B}
  \textbf{\bibinfo{volume}{72}}, \bibinfo{pages}{155111}
  (\bibinfo{year}{2005}).

\bibitem[{\citenamefont{Mizokawa et~al.}(2000)\citenamefont{Mizokawa, Khomskii,
  Sawatzky, and Johnston}}]{Mizokawa:2000wq}
\bibinfo{author}{\bibfnamefont{T.}~\bibnamefont{Mizokawa}},
  \bibinfo{author}{\bibfnamefont{D.~I.} \bibnamefont{Khomskii}},
  \bibinfo{author}{\bibfnamefont{G.}~\bibnamefont{Sawatzky}}, \bibnamefont{and}
  \bibinfo{author}{\bibfnamefont{S.}~\bibnamefont{Johnston}},
  \bibinfo{journal}{Physical review B. Condensed matter and materials physics}
  \textbf{\bibinfo{volume}{61}}, \bibinfo{pages}{11263} (\bibinfo{year}{2000}).

\bibitem[{\citenamefont{Mazin et~al.}(2007)\citenamefont{Mazin, Khomskii,
  Lengsdorf, Alonso, Marshall, Ibberson, Podlesnyak, Martinez-Lope,
  Abd-Elmeguid, and Johnston}}]{Mazin:2007jx}
\bibinfo{author}{\bibfnamefont{I.~I.} \bibnamefont{Mazin}},
  \bibinfo{author}{\bibfnamefont{D.~I.} \bibnamefont{Khomskii}},
  \bibinfo{author}{\bibfnamefont{R.}~\bibnamefont{Lengsdorf}},
  \bibinfo{author}{\bibfnamefont{J.}~\bibnamefont{Alonso}},
  \bibinfo{author}{\bibfnamefont{W.~G.} \bibnamefont{Marshall}},
  \bibinfo{author}{\bibfnamefont{R.~A.} \bibnamefont{Ibberson}},
  \bibinfo{author}{\bibfnamefont{A.}~\bibnamefont{Podlesnyak}},
  \bibinfo{author}{\bibfnamefont{M.~J.} \bibnamefont{Martinez-Lope}},
  \bibinfo{author}{\bibfnamefont{M.~M.} \bibnamefont{Abd-Elmeguid}},
  \bibnamefont{and} \bibinfo{author}{\bibfnamefont{S.}~\bibnamefont{Johnston}},
  \bibinfo{journal}{Physical Review Letters} \textbf{\bibinfo{volume}{98}},
  \bibinfo{pages}{176406} (\bibinfo{year}{2007}).

\bibitem[{\citenamefont{Park et~al.}(2012)\citenamefont{Park, Millis,
  Marianetti, and Johnston}}]{Park:2012hg}
\bibinfo{author}{\bibfnamefont{H.}~\bibnamefont{Park}},
  \bibinfo{author}{\bibfnamefont{A.}~\bibnamefont{Millis}},
  \bibinfo{author}{\bibfnamefont{C.~A.} \bibnamefont{Marianetti}},
  \bibnamefont{and} \bibinfo{author}{\bibfnamefont{S.}~\bibnamefont{Johnston}},
  \bibinfo{journal}{Physical Review Letters} \textbf{\bibinfo{volume}{109}},
  \bibinfo{pages}{156402} (\bibinfo{year}{2012}).

\bibitem[{\citenamefont{Johnston et~al.}(2014)\citenamefont{Johnston,
  Mukherjee, Elfimov, Bericu, and Sawatzky}}]{Johnston:2014tv}
\bibinfo{author}{\bibfnamefont{S.}~\bibnamefont{Johnston}},
  \bibinfo{author}{\bibfnamefont{A.}~\bibnamefont{Mukherjee}},
  \bibinfo{author}{\bibfnamefont{I.}~\bibnamefont{Elfimov}},
  \bibinfo{author}{\bibfnamefont{M.}~\bibnamefont{Bericu}}, \bibnamefont{and}
  \bibinfo{author}{\bibfnamefont{G.~A.} \bibnamefont{Sawatzky}},
  \bibinfo{journal}{Physical Review Letters} \textbf{\bibinfo{volume}{112}},
  \bibinfo{pages}{106404} (\bibinfo{year}{2014}).

\bibitem[{\citenamefont{Garc{\'\i}a-Mu{\~n}oz
  et~al.}(1992)\citenamefont{Garc{\'\i}a-Mu{\~n}oz, Rodr{\'\i}guez-Carvajal,
  and Lacorre}}]{GarciaMunoz:1992dj}
\bibinfo{author}{\bibfnamefont{J.~L.} \bibnamefont{Garc{\'\i}a-Mu{\~n}oz}},
  \bibinfo{author}{\bibfnamefont{J.}~\bibnamefont{Rodr{\'\i}guez-Carvajal}},
  \bibnamefont{and} \bibinfo{author}{\bibfnamefont{P.}~\bibnamefont{Lacorre}},
  \bibinfo{journal}{EPL (Europhysics Letters)} \textbf{\bibinfo{volume}{20}},
  \bibinfo{pages}{241} (\bibinfo{year}{1992}).

\bibitem[{\citenamefont{Scagnoli et~al.}(2006)\citenamefont{Scagnoli, Staub,
  Mulders, Janousch, Meijer, Hammerl, Tonnerre, and Stojic}}]{Scagnoli:2006ja}
\bibinfo{author}{\bibfnamefont{V.}~\bibnamefont{Scagnoli}},
  \bibinfo{author}{\bibfnamefont{U.}~\bibnamefont{Staub}},
  \bibinfo{author}{\bibfnamefont{A.}~\bibnamefont{Mulders}},
  \bibinfo{author}{\bibfnamefont{M.}~\bibnamefont{Janousch}},
  \bibinfo{author}{\bibfnamefont{G.}~\bibnamefont{Meijer}},
  \bibinfo{author}{\bibfnamefont{G.}~\bibnamefont{Hammerl}},
  \bibinfo{author}{\bibfnamefont{J.}~\bibnamefont{Tonnerre}}, \bibnamefont{and}
  \bibinfo{author}{\bibfnamefont{N.}~\bibnamefont{Stojic}},
  \bibinfo{journal}{Physical Review B} \textbf{\bibinfo{volume}{73}},
  \bibinfo{pages}{100409} (\bibinfo{year}{2006}).

\bibitem[{\citenamefont{Scagnoli et~al.}(2008)\citenamefont{Scagnoli, Staub,
  Bodenthin, Garcia-Fernandez, Mulders, Meijer, and Hammerl}}]{Scagnoli:2008iu}
\bibinfo{author}{\bibfnamefont{V.}~\bibnamefont{Scagnoli}},
  \bibinfo{author}{\bibfnamefont{U.}~\bibnamefont{Staub}},
  \bibinfo{author}{\bibfnamefont{Y.}~\bibnamefont{Bodenthin}},
  \bibinfo{author}{\bibfnamefont{M.}~\bibnamefont{Garcia-Fernandez}},
  \bibinfo{author}{\bibfnamefont{A.~M.} \bibnamefont{Mulders}},
  \bibinfo{author}{\bibfnamefont{G.~I.} \bibnamefont{Meijer}},
  \bibnamefont{and} \bibinfo{author}{\bibfnamefont{G.}~\bibnamefont{Hammerl}},
  \bibinfo{journal}{Physical Review B} \textbf{\bibinfo{volume}{77}},
  \bibinfo{pages}{115138} (\bibinfo{year}{2008}).

\bibitem[{\citenamefont{Staub et~al.}(2007)\citenamefont{Staub,
  Garc{\'\i}a-Fern{\'a}ndez, Mulders, Bodenthin, Mart{\'\i}nez-Lope, and
  Alonso}}]{Staub:2007gr}
\bibinfo{author}{\bibfnamefont{U.}~\bibnamefont{Staub}},
  \bibinfo{author}{\bibfnamefont{M.}~\bibnamefont{Garc{\'\i}a-Fern{\'a}ndez}},
  \bibinfo{author}{\bibfnamefont{A.~M.} \bibnamefont{Mulders}},
  \bibinfo{author}{\bibfnamefont{Y.}~\bibnamefont{Bodenthin}},
  \bibinfo{author}{\bibfnamefont{M.~J.} \bibnamefont{Mart{\'\i}nez-Lope}},
  \bibnamefont{and} \bibinfo{author}{\bibfnamefont{J.~A.}
  \bibnamefont{Alonso}}, \bibinfo{journal}{JOURNAL OF PHYSICS CONDENSED MATTER}
  \textbf{\bibinfo{volume}{19}}, \bibinfo{pages}{092201}
  (\bibinfo{year}{2007}).

\bibitem[{\citenamefont{Bodenthin et~al.}(2011)\citenamefont{Bodenthin, Staub,
  Piamonteze, Garcia-Fernandez, Martinez-Lope, and Alonso}}]{Bodenthin:2011bn}
\bibinfo{author}{\bibfnamefont{Y.}~\bibnamefont{Bodenthin}},
  \bibinfo{author}{\bibfnamefont{U.}~\bibnamefont{Staub}},
  \bibinfo{author}{\bibfnamefont{C.}~\bibnamefont{Piamonteze}},
  \bibinfo{author}{\bibfnamefont{M.}~\bibnamefont{Garcia-Fernandez}},
  \bibinfo{author}{\bibfnamefont{M.~J.} \bibnamefont{Martinez-Lope}},
  \bibnamefont{and} \bibinfo{author}{\bibfnamefont{J.~A.}
  \bibnamefont{Alonso}}, \bibinfo{journal}{JOURNAL OF PHYSICS CONDENSED MATTER}
  \textbf{\bibinfo{volume}{23}}, \bibinfo{pages}{036002}
  (\bibinfo{year}{2011}).

\bibitem[{\citenamefont{Zhou et~al.}(2003)\citenamefont{Zhou, Goodenough, and
  Dabrowski}}]{Zhou:2003bf}
\bibinfo{author}{\bibfnamefont{J.}~\bibnamefont{Zhou}},
  \bibinfo{author}{\bibfnamefont{J.}~\bibnamefont{Goodenough}},
  \bibnamefont{and}
  \bibinfo{author}{\bibfnamefont{B.}~\bibnamefont{Dabrowski}},
  \bibinfo{journal}{Physical Review B} \textbf{\bibinfo{volume}{67}},
  \bibinfo{pages}{020404} (\bibinfo{year}{2003}).

\bibitem[{\citenamefont{Kumar et~al.}(2013)\citenamefont{Kumar, Rajeev, Alonso,
  and Mart{\'\i}nez-Lope}}]{Kumar:2013be}
\bibinfo{author}{\bibfnamefont{D.}~\bibnamefont{Kumar}},
  \bibinfo{author}{\bibfnamefont{K.~P.} \bibnamefont{Rajeev}},
  \bibinfo{author}{\bibfnamefont{J.~A.} \bibnamefont{Alonso}},
  \bibnamefont{and} \bibinfo{author}{\bibfnamefont{M.~J.}
  \bibnamefont{Mart{\'\i}nez-Lope}}, \bibinfo{journal}{Physical Review B}
  \textbf{\bibinfo{volume}{88}}, \bibinfo{pages}{014410}
  (\bibinfo{year}{2013}).

\bibitem[{\citenamefont{Zhou and Goodenough}(2014)}]{Zhou:2014er}
\bibinfo{author}{\bibfnamefont{J.~S.} \bibnamefont{Zhou}} \bibnamefont{and}
  \bibinfo{author}{\bibfnamefont{J.}~\bibnamefont{Goodenough}},
  \bibinfo{journal}{Physical review B. Condensed matter and materials physics}
  \textbf{\bibinfo{volume}{89}}, \bibinfo{pages}{245138}
  (\bibinfo{year}{2014}).

\bibitem[{\citenamefont{Sarma et~al.}(1994)\citenamefont{Sarma, Shanthi, and
  Mahadevan}}]{Sarma:1994eo}
\bibinfo{author}{\bibfnamefont{D.~D.} \bibnamefont{Sarma}},
  \bibinfo{author}{\bibfnamefont{N.}~\bibnamefont{Shanthi}}, \bibnamefont{and}
  \bibinfo{author}{\bibfnamefont{P.}~\bibnamefont{Mahadevan}},
  \bibinfo{journal}{JOURNAL OF PHYSICS CONDENSED MATTER}
  \textbf{\bibinfo{volume}{6}}, \bibinfo{pages}{10467} (\bibinfo{year}{1994}).

\bibitem[{\citenamefont{Zhou et~al.}(2004)\citenamefont{Zhou, Goodenough, and
  Dabrowski}}]{Zhou:2004fv}
\bibinfo{author}{\bibfnamefont{J.}~\bibnamefont{Zhou}},
  \bibinfo{author}{\bibfnamefont{J.}~\bibnamefont{Goodenough}},
  \bibnamefont{and}
  \bibinfo{author}{\bibfnamefont{B.}~\bibnamefont{Dabrowski}},
  \bibinfo{journal}{Physical Review B} \textbf{\bibinfo{volume}{70}},
  \bibinfo{pages}{081102} (\bibinfo{year}{2004}).

\bibitem[{\citenamefont{Barman et~al.}(1994)\citenamefont{Barman, Chainani, and
  Sarma}}]{Barman:1994ee}
\bibinfo{author}{\bibfnamefont{S.}~\bibnamefont{Barman}},
  \bibinfo{author}{\bibfnamefont{A.}~\bibnamefont{Chainani}}, \bibnamefont{and}
  \bibinfo{author}{\bibfnamefont{D.}~\bibnamefont{Sarma}},
  \bibinfo{journal}{Physical Review B} \textbf{\bibinfo{volume}{49}},
  \bibinfo{pages}{8475} (\bibinfo{year}{1994}).

\bibitem[{\citenamefont{Zaanen et~al.}(1985)\citenamefont{Zaanen, Sawatzky, and
  Allen}}]{Zaanen:1985wp}
\bibinfo{author}{\bibfnamefont{J.}~\bibnamefont{Zaanen}},
  \bibinfo{author}{\bibfnamefont{G.}~\bibnamefont{Sawatzky}}, \bibnamefont{and}
  \bibinfo{author}{\bibfnamefont{J.}~\bibnamefont{Allen}},
  \bibinfo{journal}{Physical Review Letters} \textbf{\bibinfo{volume}{55}},
  \bibinfo{pages}{418} (\bibinfo{year}{1985}).

\bibitem[{\citenamefont{Medarde et~al.}(1997)\citenamefont{Medarde, Purdie,
  Grioni, Hengsberger, Baer, and Lacorre}}]{Medarde:1997bu}
\bibinfo{author}{\bibfnamefont{M.}~\bibnamefont{Medarde}},
  \bibinfo{author}{\bibfnamefont{D.}~\bibnamefont{Purdie}},
  \bibinfo{author}{\bibfnamefont{M.}~\bibnamefont{Grioni}},
  \bibinfo{author}{\bibfnamefont{M.}~\bibnamefont{Hengsberger}},
  \bibinfo{author}{\bibfnamefont{Y.}~\bibnamefont{Baer}}, \bibnamefont{and}
  \bibinfo{author}{\bibfnamefont{P.}~\bibnamefont{Lacorre}},
  \bibinfo{journal}{EPL (Europhysics Letters)} \textbf{\bibinfo{volume}{37}},
  \bibinfo{pages}{483} (\bibinfo{year}{1997}).

\bibitem[{\citenamefont{Vobornik et~al.}(1999)\citenamefont{Vobornik, Perfetti,
  Zacchigna, Grioni, Margaritondo, Mesot, Medarde, and
  Lacorre}}]{Vobornik:1999de}
\bibinfo{author}{\bibfnamefont{I.}~\bibnamefont{Vobornik}},
  \bibinfo{author}{\bibfnamefont{L.}~\bibnamefont{Perfetti}},
  \bibinfo{author}{\bibfnamefont{M.}~\bibnamefont{Zacchigna}},
  \bibinfo{author}{\bibfnamefont{M.}~\bibnamefont{Grioni}},
  \bibinfo{author}{\bibfnamefont{G.}~\bibnamefont{Margaritondo}},
  \bibinfo{author}{\bibfnamefont{J.}~\bibnamefont{Mesot}},
  \bibinfo{author}{\bibfnamefont{M.}~\bibnamefont{Medarde}}, \bibnamefont{and}
  \bibinfo{author}{\bibfnamefont{P.}~\bibnamefont{Lacorre}},
  \bibinfo{journal}{Physical Review B} \textbf{\bibinfo{volume}{60}},
  \bibinfo{pages}{R8426} (\bibinfo{year}{1999}).

\bibitem[{\citenamefont{Okazaki et~al.}(2003)\citenamefont{Okazaki, Mizokawa,
  and Fujimori}}]{Okazaki:2003to}
\bibinfo{author}{\bibfnamefont{K.}~\bibnamefont{Okazaki}},
  \bibinfo{author}{\bibfnamefont{T.}~\bibnamefont{Mizokawa}}, \bibnamefont{and}
  \bibinfo{author}{\bibfnamefont{A.}~\bibnamefont{Fujimori}},
  \bibinfo{journal}{Physical Review B} \textbf{\bibinfo{volume}{67}},
  \bibinfo{pages}{73101} (\bibinfo{year}{2003}).

\bibitem[{\citenamefont{Schwier et~al.}(2012)\citenamefont{Schwier, Scherwitzl,
  Vydrov{\`a}, Garcia-Fernandez, Gibert, Zubko, Garnier, Triscone, and
  Aebi}}]{Schwier:2012gqa}
\bibinfo{author}{\bibfnamefont{E.~F.} \bibnamefont{Schwier}},
  \bibinfo{author}{\bibfnamefont{R.}~\bibnamefont{Scherwitzl}},
  \bibinfo{author}{\bibfnamefont{Z.}~\bibnamefont{Vydrov{\`a}}},
  \bibinfo{author}{\bibfnamefont{M.}~\bibnamefont{Garcia-Fernandez}},
  \bibinfo{author}{\bibfnamefont{M.}~\bibnamefont{Gibert}},
  \bibinfo{author}{\bibfnamefont{P.}~\bibnamefont{Zubko}},
  \bibinfo{author}{\bibfnamefont{M.~G.} \bibnamefont{Garnier}},
  \bibinfo{author}{\bibfnamefont{J.~M.} \bibnamefont{Triscone}},
  \bibnamefont{and} \bibinfo{author}{\bibfnamefont{P.}~\bibnamefont{Aebi}},
  \bibinfo{journal}{Physical Review B} \textbf{\bibinfo{volume}{86}},
  \bibinfo{pages}{195147} (\bibinfo{year}{2012}).

\bibitem[{\citenamefont{Katsufuji et~al.}(1995)\citenamefont{Katsufuji,
  Okimoto, Arima, Tokura, and Torrance}}]{Katsufuji:1995dy}
\bibinfo{author}{\bibfnamefont{T.}~\bibnamefont{Katsufuji}},
  \bibinfo{author}{\bibfnamefont{Y.}~\bibnamefont{Okimoto}},
  \bibinfo{author}{\bibfnamefont{T.}~\bibnamefont{Arima}},
  \bibinfo{author}{\bibfnamefont{Y.}~\bibnamefont{Tokura}}, \bibnamefont{and}
  \bibinfo{author}{\bibfnamefont{J.~B.} \bibnamefont{Torrance}},
  \bibinfo{journal}{Physical Review B} \textbf{\bibinfo{volume}{51}},
  \bibinfo{pages}{4830} (\bibinfo{year}{1995}).

\bibitem[{\citenamefont{Mizokawa et~al.}(1995)\citenamefont{Mizokawa, Fujimori,
  Arima, Tokura, and Mori}}]{Mizokawa:1995vq}
\bibinfo{author}{\bibfnamefont{T.}~\bibnamefont{Mizokawa}},
  \bibinfo{author}{\bibfnamefont{A.}~\bibnamefont{Fujimori}},
  \bibinfo{author}{\bibfnamefont{T.}~\bibnamefont{Arima}},
  \bibinfo{author}{\bibfnamefont{Y.}~\bibnamefont{Tokura}}, \bibnamefont{and}
  \bibinfo{author}{\bibfnamefont{N.}~\bibnamefont{Mori}},
  \bibinfo{journal}{Physical Review-Section B-Condensed Matter}
  \textbf{\bibinfo{volume}{52}}, \bibinfo{pages}{13865} (\bibinfo{year}{1995}).

\bibitem[{\citenamefont{Lee et~al.}(2011)\citenamefont{Lee, Chen, and
  Balents}}]{Lee:2011bq}
\bibinfo{author}{\bibfnamefont{S.}~\bibnamefont{Lee}},
  \bibinfo{author}{\bibfnamefont{R.}~\bibnamefont{Chen}}, \bibnamefont{and}
  \bibinfo{author}{\bibfnamefont{L.}~\bibnamefont{Balents}},
  \bibinfo{journal}{Physical Review Letters} \textbf{\bibinfo{volume}{106}},
  \bibinfo{pages}{016405} (\bibinfo{year}{2011}).

\bibitem[{\citenamefont{Subedi et~al.}(2015)\citenamefont{Subedi, Peil, and
  Georges}}]{Subedi:2015en}
\bibinfo{author}{\bibfnamefont{A.}~\bibnamefont{Subedi}},
  \bibinfo{author}{\bibfnamefont{O.~E.} \bibnamefont{Peil}}, \bibnamefont{and}
  \bibinfo{author}{\bibfnamefont{A.}~\bibnamefont{Georges}},
  \bibinfo{journal}{Physical Review B} \textbf{\bibinfo{volume}{91}},
  \bibinfo{pages}{75128} (\bibinfo{year}{2015}).

\bibitem[{\citenamefont{Piamonteze
  et~al.}(2005{\natexlab{a}})\citenamefont{Piamonteze, de~Groot, and
  Tolentino}}]{Piamonteze:2005ux}
\bibinfo{author}{\bibfnamefont{C.}~\bibnamefont{Piamonteze}},
  \bibinfo{author}{\bibfnamefont{F.}~\bibnamefont{de~Groot}}, \bibnamefont{and}
  \bibinfo{author}{\bibfnamefont{H.}~\bibnamefont{Tolentino}},
  \bibinfo{journal}{Physical Review B} \textbf{\bibinfo{volume}{71}},
  \bibinfo{pages}{020406} (\bibinfo{year}{2005}{\natexlab{a}}).

\bibitem[{\citenamefont{van~der Laan and Figueroa}(2014)}]{vanderLaan:2014if}
\bibinfo{author}{\bibfnamefont{G.}~\bibnamefont{van~der Laan}}
  \bibnamefont{and} \bibinfo{author}{\bibfnamefont{A.~I.}
  \bibnamefont{Figueroa}}, \bibinfo{journal}{Coordination Chemistry Reviews}
  \textbf{\bibinfo{volume}{277-278}}, \bibinfo{pages}{95}
  (\bibinfo{year}{2014}).

\bibitem[{\citenamefont{Liu et~al.}(2011)\citenamefont{Liu, Okamoto,
  Van~Veenendaal, Kareev, Gray, Ryan, Freeland, and Chakhalian}}]{Liu:2011ej}
\bibinfo{author}{\bibfnamefont{J.}~\bibnamefont{Liu}},
  \bibinfo{author}{\bibfnamefont{S.}~\bibnamefont{Okamoto}},
  \bibinfo{author}{\bibfnamefont{M.}~\bibnamefont{Van~Veenendaal}},
  \bibinfo{author}{\bibfnamefont{M.}~\bibnamefont{Kareev}},
  \bibinfo{author}{\bibfnamefont{B.}~\bibnamefont{Gray}},
  \bibinfo{author}{\bibfnamefont{P.}~\bibnamefont{Ryan}},
  \bibinfo{author}{\bibfnamefont{J.}~\bibnamefont{Freeland}}, \bibnamefont{and}
  \bibinfo{author}{\bibfnamefont{J.}~\bibnamefont{Chakhalian}},
  \bibinfo{journal}{Physical Review B} \textbf{\bibinfo{volume}{83}},
  \bibinfo{pages}{161102} (\bibinfo{year}{2011}).

\bibitem[{\citenamefont{Gou et~al.}(2011)\citenamefont{Gou, Grinberg, Rappe,
  and Rondinelli}}]{Gou:2011ky}
\bibinfo{author}{\bibfnamefont{G.}~\bibnamefont{Gou}},
  \bibinfo{author}{\bibfnamefont{I.}~\bibnamefont{Grinberg}},
  \bibinfo{author}{\bibfnamefont{A.~M.} \bibnamefont{Rappe}}, \bibnamefont{and}
  \bibinfo{author}{\bibfnamefont{J.~M.} \bibnamefont{Rondinelli}},
  \bibinfo{journal}{Physical Review B} \textbf{\bibinfo{volume}{84}},
  \bibinfo{pages}{144101} (\bibinfo{year}{2011}).

\bibitem[{\citenamefont{Medarde et~al.}(1992)\citenamefont{Medarde, Fontaine,
  Garc{\'\i}a-Mu{\~n}oz, Rodr{\'\i}guez-Carvajal, De~Santis, Sacchi, Rossi, and
  Lacorre}}]{Medarde:1992ds}
\bibinfo{author}{\bibfnamefont{M.}~\bibnamefont{Medarde}},
  \bibinfo{author}{\bibfnamefont{A.}~\bibnamefont{Fontaine}},
  \bibinfo{author}{\bibfnamefont{J.}~\bibnamefont{Garc{\'\i}a-Mu{\~n}oz}},
  \bibinfo{author}{\bibfnamefont{J.}~\bibnamefont{Rodr{\'\i}guez-Carvajal}},
  \bibinfo{author}{\bibfnamefont{M.}~\bibnamefont{De~Santis}},
  \bibinfo{author}{\bibfnamefont{M.}~\bibnamefont{Sacchi}},
  \bibinfo{author}{\bibfnamefont{G.}~\bibnamefont{Rossi}}, \bibnamefont{and}
  \bibinfo{author}{\bibfnamefont{P.}~\bibnamefont{Lacorre}},
  \bibinfo{journal}{Physical Review B} \textbf{\bibinfo{volume}{46}},
  \bibinfo{pages}{14975} (\bibinfo{year}{1992}).

\bibitem[{\citenamefont{Abbate et~al.}(2002)\citenamefont{Abbate, Zampieri,
  Prado, and Caneiro}}]{Abbate:2002ti}
\bibinfo{author}{\bibfnamefont{M.}~\bibnamefont{Abbate}},
  \bibinfo{author}{\bibfnamefont{G.}~\bibnamefont{Zampieri}},
  \bibinfo{author}{\bibfnamefont{F.}~\bibnamefont{Prado}}, \bibnamefont{and}
  \bibinfo{author}{\bibfnamefont{A.}~\bibnamefont{Caneiro}},
  \bibinfo{journal}{Physical Review B} \textbf{\bibinfo{volume}{65}},
  \bibinfo{pages}{155101} (\bibinfo{year}{2002}).

\bibitem[{\citenamefont{Suntivich et~al.}(2014)\citenamefont{Suntivich, Hong,
  Lee, Rondinelli, Yang, Goodenough, Dabrowski, Freeland, and
  Shao-Horn}}]{Suntivich:2014ft}
\bibinfo{author}{\bibfnamefont{J.}~\bibnamefont{Suntivich}},
  \bibinfo{author}{\bibfnamefont{W.~T.} \bibnamefont{Hong}},
  \bibinfo{author}{\bibfnamefont{Y.-L.} \bibnamefont{Lee}},
  \bibinfo{author}{\bibfnamefont{J.~M.} \bibnamefont{Rondinelli}},
  \bibinfo{author}{\bibfnamefont{W.}~\bibnamefont{Yang}},
  \bibinfo{author}{\bibfnamefont{J.~B.} \bibnamefont{Goodenough}},
  \bibinfo{author}{\bibfnamefont{B.}~\bibnamefont{Dabrowski}},
  \bibinfo{author}{\bibfnamefont{J.~W.} \bibnamefont{Freeland}},
  \bibnamefont{and}
  \bibinfo{author}{\bibfnamefont{Y.}~\bibnamefont{Shao-Horn}},
  \bibinfo{journal}{Journal Of Physical Chemistry C}
  \textbf{\bibinfo{volume}{118}}, \bibinfo{pages}{1856} (\bibinfo{year}{2014}).

\bibitem[{\citenamefont{Sarma et~al.}(1996)\citenamefont{Sarma, Shanthi, and
  Mahadevan}}]{Sarma:1996bs}
\bibinfo{author}{\bibfnamefont{D.}~\bibnamefont{Sarma}},
  \bibinfo{author}{\bibfnamefont{N.}~\bibnamefont{Shanthi}}, \bibnamefont{and}
  \bibinfo{author}{\bibfnamefont{P.}~\bibnamefont{Mahadevan}},
  \bibinfo{journal}{Physical Review B} \textbf{\bibinfo{volume}{54}},
  \bibinfo{pages}{1622} (\bibinfo{year}{1996}).

\bibitem[{\citenamefont{de~Groot}(2005)}]{deGroot:2005wg}
\bibinfo{author}{\bibfnamefont{F.}~\bibnamefont{de~Groot}},
  \bibinfo{journal}{Coordination Chemistry Reviews}
  \textbf{\bibinfo{volume}{249}}, \bibinfo{pages}{31} (\bibinfo{year}{2005}).

\bibitem[{\citenamefont{Medling et~al.}(2012)\citenamefont{Medling, Lee, Zheng,
  Mitchell, Freeland, Harmon, and Bridges}}]{Medling:2012wm}
\bibinfo{author}{\bibfnamefont{S.}~\bibnamefont{Medling}},
  \bibinfo{author}{\bibfnamefont{Y.}~\bibnamefont{Lee}},
  \bibinfo{author}{\bibfnamefont{H.}~\bibnamefont{Zheng}},
  \bibinfo{author}{\bibfnamefont{J.~F.} \bibnamefont{Mitchell}},
  \bibinfo{author}{\bibfnamefont{J.~W.} \bibnamefont{Freeland}},
  \bibinfo{author}{\bibfnamefont{B.~N.} \bibnamefont{Harmon}},
  \bibnamefont{and} \bibinfo{author}{\bibfnamefont{F.}~\bibnamefont{Bridges}},
  \bibinfo{journal}{Physical Review Letters} \textbf{\bibinfo{volume}{109}},
  \bibinfo{pages}{157204} (\bibinfo{year}{2012}).

\bibitem[{\citenamefont{Fujimori and Minami}(1984)}]{Fujimori:1984ju}
\bibinfo{author}{\bibfnamefont{A.}~\bibnamefont{Fujimori}} \bibnamefont{and}
  \bibinfo{author}{\bibfnamefont{F.}~\bibnamefont{Minami}},
  \bibinfo{journal}{Physical Review B} \textbf{\bibinfo{volume}{30}},
  \bibinfo{pages}{957} (\bibinfo{year}{1984}).

\bibitem[{\citenamefont{Kuiper et~al.}(1989)\citenamefont{Kuiper, Kruizinga,
  Ghijsen, Sawatzky, and Verweij}}]{Kuiper:1989ep}
\bibinfo{author}{\bibfnamefont{P.}~\bibnamefont{Kuiper}},
  \bibinfo{author}{\bibfnamefont{G.}~\bibnamefont{Kruizinga}},
  \bibinfo{author}{\bibfnamefont{J.}~\bibnamefont{Ghijsen}},
  \bibinfo{author}{\bibfnamefont{G.}~\bibnamefont{Sawatzky}}, \bibnamefont{and}
  \bibinfo{author}{\bibfnamefont{H.}~\bibnamefont{Verweij}},
  \bibinfo{journal}{Physical Review Letters} \textbf{\bibinfo{volume}{62}},
  \bibinfo{pages}{221} (\bibinfo{year}{1989}).

\bibitem[{\citenamefont{Van~Elp et~al.}(1992)\citenamefont{Van~Elp, Eskes,
  Kuiper, and Sawatzky}}]{VanElp:1992ko}
\bibinfo{author}{\bibfnamefont{J.}~\bibnamefont{Van~Elp}},
  \bibinfo{author}{\bibfnamefont{H.}~\bibnamefont{Eskes}},
  \bibinfo{author}{\bibfnamefont{P.}~\bibnamefont{Kuiper}}, \bibnamefont{and}
  \bibinfo{author}{\bibfnamefont{G.}~\bibnamefont{Sawatzky}},
  \bibinfo{journal}{Physical Review B} \textbf{\bibinfo{volume}{45}},
  \bibinfo{pages}{1612} (\bibinfo{year}{1992}).

\bibitem[{\citenamefont{Van~Veenendaal and
  Sawatzky}(1993)}]{VanVeenendaal:1993dx}
\bibinfo{author}{\bibfnamefont{M.}~\bibnamefont{Van~Veenendaal}}
  \bibnamefont{and} \bibinfo{author}{\bibfnamefont{G.}~\bibnamefont{Sawatzky}},
  \bibinfo{journal}{Physical Review Letters} \textbf{\bibinfo{volume}{70}},
  \bibinfo{pages}{2459} (\bibinfo{year}{1993}).

\bibitem[{\citenamefont{Sawatzky and Allen}(1984)}]{Sawatzky:1984jt}
\bibinfo{author}{\bibfnamefont{G.}~\bibnamefont{Sawatzky}} \bibnamefont{and}
  \bibinfo{author}{\bibfnamefont{J.}~\bibnamefont{Allen}},
  \bibinfo{journal}{Physical Review Letters} \textbf{\bibinfo{volume}{53}},
  \bibinfo{pages}{2339} (\bibinfo{year}{1984}).

\bibitem[{\citenamefont{Zhang and Rice}(1988)}]{Zhang:1988jf}
\bibinfo{author}{\bibfnamefont{F.}~\bibnamefont{Zhang}} \bibnamefont{and}
  \bibinfo{author}{\bibfnamefont{T.}~\bibnamefont{Rice}},
  \bibinfo{journal}{Physical Review B} \textbf{\bibinfo{volume}{37}},
  \bibinfo{pages}{3759} (\bibinfo{year}{1988}).

\bibitem[{\citenamefont{Eskes and Sawatzky}(1988)}]{Eskes:1988ef}
\bibinfo{author}{\bibfnamefont{H.}~\bibnamefont{Eskes}} \bibnamefont{and}
  \bibinfo{author}{\bibfnamefont{G.}~\bibnamefont{Sawatzky}},
  \bibinfo{journal}{Physical Review Letters} \textbf{\bibinfo{volume}{61}},
  \bibinfo{pages}{1415} (\bibinfo{year}{1988}).

\bibitem[{\citenamefont{Van~Veenendaal and
  Sawatzky}(1994)}]{VanVeenendaal:1994kh}
\bibinfo{author}{\bibfnamefont{M.}~\bibnamefont{Van~Veenendaal}}
  \bibnamefont{and} \bibinfo{author}{\bibfnamefont{G.}~\bibnamefont{Sawatzky}},
  \bibinfo{journal}{Physical Review B} \textbf{\bibinfo{volume}{50}},
  \bibinfo{pages}{11326} (\bibinfo{year}{1994}).

\bibitem[{\citenamefont{Gordon et~al.}(2008)\citenamefont{Gordon, Seidler,
  Fister, Haverkort, Sawatzky, Tanaka, and Sham}}]{Gordon:2008hb}
\bibinfo{author}{\bibfnamefont{R.~A.} \bibnamefont{Gordon}},
  \bibinfo{author}{\bibfnamefont{G.~T.} \bibnamefont{Seidler}},
  \bibinfo{author}{\bibfnamefont{T.~T.} \bibnamefont{Fister}},
  \bibinfo{author}{\bibfnamefont{M.~W.} \bibnamefont{Haverkort}},
  \bibinfo{author}{\bibfnamefont{G.~A.} \bibnamefont{Sawatzky}},
  \bibinfo{author}{\bibfnamefont{A.}~\bibnamefont{Tanaka}}, \bibnamefont{and}
  \bibinfo{author}{\bibfnamefont{T.~K.} \bibnamefont{Sham}},
  \bibinfo{journal}{EPL (Europhysics Letters)} \textbf{\bibinfo{volume}{81}},
  \bibinfo{pages}{26004} (\bibinfo{year}{2008}).

\bibitem[{\citenamefont{Van Der~Laan et~al.}(1986)\citenamefont{Van Der~Laan,
  Zaanen, Sawatzky, Karnatak, and Esteva}}]{VanDerLaan:1986ez}
\bibinfo{author}{\bibfnamefont{G.}~\bibnamefont{Van Der~Laan}},
  \bibinfo{author}{\bibfnamefont{J.}~\bibnamefont{Zaanen}},
  \bibinfo{author}{\bibfnamefont{G.}~\bibnamefont{Sawatzky}},
  \bibinfo{author}{\bibfnamefont{R.}~\bibnamefont{Karnatak}}, \bibnamefont{and}
  \bibinfo{author}{\bibfnamefont{J.-M.} \bibnamefont{Esteva}},
  \bibinfo{journal}{Physical Review B} \textbf{\bibinfo{volume}{33}},
  \bibinfo{pages}{4253} (\bibinfo{year}{1986}).

\bibitem[{\citenamefont{Ushakov et~al.}(2011)\citenamefont{Ushakov, Streltsov,
  and Khomskii}}]{Ushakov:2011fy}
\bibinfo{author}{\bibfnamefont{A.~V.} \bibnamefont{Ushakov}},
  \bibinfo{author}{\bibfnamefont{S.~V.} \bibnamefont{Streltsov}},
  \bibnamefont{and} \bibinfo{author}{\bibfnamefont{D.~I.}
  \bibnamefont{Khomskii}}, \bibinfo{journal}{JOURNAL OF PHYSICS CONDENSED
  MATTER} \textbf{\bibinfo{volume}{23}}, \bibinfo{pages}{445601}
  (\bibinfo{year}{2011}).

\bibitem[{\citenamefont{Ohta et~al.}(1991)\citenamefont{Ohta, Tohyama, and
  Maekawa}}]{Ohta:1991ft}
\bibinfo{author}{\bibfnamefont{Y.}~\bibnamefont{Ohta}},
  \bibinfo{author}{\bibfnamefont{T.}~\bibnamefont{Tohyama}}, \bibnamefont{and}
  \bibinfo{author}{\bibfnamefont{S.}~\bibnamefont{Maekawa}},
  \bibinfo{journal}{Physical Review B} \textbf{\bibinfo{volume}{43}},
  \bibinfo{pages}{2968} (\bibinfo{year}{1991}).

\bibitem[{\citenamefont{Cavalleri et~al.}(2004)\citenamefont{Cavalleri, Chong,
  Fourmaux, Glover, Heimann, Kieffer, Mun, Padmore, and
  Schoenlein}}]{Cavalleri:2004kf}
\bibinfo{author}{\bibfnamefont{A.}~\bibnamefont{Cavalleri}},
  \bibinfo{author}{\bibfnamefont{H.~H.~W.} \bibnamefont{Chong}},
  \bibinfo{author}{\bibfnamefont{S.}~\bibnamefont{Fourmaux}},
  \bibinfo{author}{\bibfnamefont{T.~E.} \bibnamefont{Glover}},
  \bibinfo{author}{\bibfnamefont{P.~A.} \bibnamefont{Heimann}},
  \bibinfo{author}{\bibfnamefont{J.~C.} \bibnamefont{Kieffer}},
  \bibinfo{author}{\bibfnamefont{B.~S.} \bibnamefont{Mun}},
  \bibinfo{author}{\bibfnamefont{H.~A.} \bibnamefont{Padmore}},
  \bibnamefont{and} \bibinfo{author}{\bibfnamefont{R.~W.}
  \bibnamefont{Schoenlein}}, \bibinfo{journal}{Physical Review B}
  \textbf{\bibinfo{volume}{69}}, \bibinfo{pages}{153106}
  (\bibinfo{year}{2004}).

\bibitem[{\citenamefont{Cavalleri et~al.}(2005)\citenamefont{Cavalleri, Rini,
  Chong, Fourmaux, Glover, Heimann, Kieffer, and
  Schoenlein}}]{Cavalleri:2005fe}
\bibinfo{author}{\bibfnamefont{A.}~\bibnamefont{Cavalleri}},
  \bibinfo{author}{\bibfnamefont{M.}~\bibnamefont{Rini}},
  \bibinfo{author}{\bibfnamefont{H.~H.~W.} \bibnamefont{Chong}},
  \bibinfo{author}{\bibfnamefont{S.}~\bibnamefont{Fourmaux}},
  \bibinfo{author}{\bibfnamefont{T.~E.} \bibnamefont{Glover}},
  \bibinfo{author}{\bibfnamefont{P.~A.} \bibnamefont{Heimann}},
  \bibinfo{author}{\bibfnamefont{J.~C.} \bibnamefont{Kieffer}},
  \bibnamefont{and} \bibinfo{author}{\bibfnamefont{R.~W.}
  \bibnamefont{Schoenlein}}, \bibinfo{journal}{Physical Review Letters}
  \textbf{\bibinfo{volume}{95}}, \bibinfo{pages}{067405}
  (\bibinfo{year}{2005}).

\bibitem[{\citenamefont{Merz et~al.}(2006)\citenamefont{Merz, Roth, Reutler,
  B{\"u}chner, Arena, Dvorak, Idzerda, Tokumitsu, and Schuppler}}]{Merz:2006jd}
\bibinfo{author}{\bibfnamefont{M.}~\bibnamefont{Merz}},
  \bibinfo{author}{\bibfnamefont{G.}~\bibnamefont{Roth}},
  \bibinfo{author}{\bibfnamefont{P.}~\bibnamefont{Reutler}},
  \bibinfo{author}{\bibfnamefont{B.}~\bibnamefont{B{\"u}chner}},
  \bibinfo{author}{\bibfnamefont{D.}~\bibnamefont{Arena}},
  \bibinfo{author}{\bibfnamefont{J.}~\bibnamefont{Dvorak}},
  \bibinfo{author}{\bibfnamefont{Y.~U.} \bibnamefont{Idzerda}},
  \bibinfo{author}{\bibfnamefont{S.}~\bibnamefont{Tokumitsu}},
  \bibnamefont{and}
  \bibinfo{author}{\bibfnamefont{S.}~\bibnamefont{Schuppler}},
  \bibinfo{journal}{Physical Review B} \textbf{\bibinfo{volume}{74}},
  \bibinfo{pages}{184414} (\bibinfo{year}{2006}).

\bibitem[{\citenamefont{Rini et~al.}(2009)\citenamefont{Rini, Zhu, Wall, Tobey,
  Ehrke, Garl, Freeland, Tomioka, Tokura, Cavalleri et~al.}}]{Rini:2009cg}
\bibinfo{author}{\bibfnamefont{M.}~\bibnamefont{Rini}},
  \bibinfo{author}{\bibfnamefont{Y.}~\bibnamefont{Zhu}},
  \bibinfo{author}{\bibfnamefont{S.}~\bibnamefont{Wall}},
  \bibinfo{author}{\bibfnamefont{R.~I.} \bibnamefont{Tobey}},
  \bibinfo{author}{\bibfnamefont{H.}~\bibnamefont{Ehrke}},
  \bibinfo{author}{\bibfnamefont{T.}~\bibnamefont{Garl}},
  \bibinfo{author}{\bibfnamefont{J.~W.} \bibnamefont{Freeland}},
  \bibinfo{author}{\bibfnamefont{Y.}~\bibnamefont{Tomioka}},
  \bibinfo{author}{\bibfnamefont{Y.}~\bibnamefont{Tokura}},
  \bibinfo{author}{\bibfnamefont{A.}~\bibnamefont{Cavalleri}},
  \bibnamefont{et~al.}, \bibinfo{journal}{Physical Review B}
  \textbf{\bibinfo{volume}{80}}, \bibinfo{pages}{155113}
  (\bibinfo{year}{2009}).

\bibitem[{\citenamefont{Wang et~al.}(2012)\citenamefont{Wang, Han, de~Medici,
  Park, Marianetti, and Millis}}]{Wang:2012dd}
\bibinfo{author}{\bibfnamefont{X.}~\bibnamefont{Wang}},
  \bibinfo{author}{\bibfnamefont{M.}~\bibnamefont{Han}},
  \bibinfo{author}{\bibfnamefont{L.}~\bibnamefont{de~Medici}},
  \bibinfo{author}{\bibfnamefont{H.}~\bibnamefont{Park}},
  \bibinfo{author}{\bibfnamefont{C.}~\bibnamefont{Marianetti}},
  \bibnamefont{and} \bibinfo{author}{\bibfnamefont{A.}~\bibnamefont{Millis}},
  \bibinfo{journal}{Physical Review B} \textbf{\bibinfo{volume}{86}},
  \bibinfo{pages}{195136} (\bibinfo{year}{2012}).

\bibitem[{\citenamefont{Piamonteze
  et~al.}(2005{\natexlab{b}})\citenamefont{Piamonteze, Tolentino, Ramos, Massa,
  Alonso, Martinez-Lope, and Casais}}]{Piamonteze:2005cd}
\bibinfo{author}{\bibfnamefont{C.}~\bibnamefont{Piamonteze}},
  \bibinfo{author}{\bibfnamefont{H.}~\bibnamefont{Tolentino}},
  \bibinfo{author}{\bibfnamefont{A.}~\bibnamefont{Ramos}},
  \bibinfo{author}{\bibfnamefont{N.}~\bibnamefont{Massa}},
  \bibinfo{author}{\bibfnamefont{J.}~\bibnamefont{Alonso}},
  \bibinfo{author}{\bibfnamefont{M.}~\bibnamefont{Martinez-Lope}},
  \bibnamefont{and} \bibinfo{author}{\bibfnamefont{M.}~\bibnamefont{Casais}},
  \bibinfo{journal}{Physical Review B} \textbf{\bibinfo{volume}{71}},
  \bibinfo{pages}{012104} (\bibinfo{year}{2005}{\natexlab{b}}).

\bibitem[{\citenamefont{Medarde et~al.}(2009)\citenamefont{Medarde, Dallera,
  Grioni, Delley, Vernay, Mesot, Sikora, Alonso, and
  Martinez-Lope}}]{Medarde:2009hg}
\bibinfo{author}{\bibfnamefont{M.}~\bibnamefont{Medarde}},
  \bibinfo{author}{\bibfnamefont{C.}~\bibnamefont{Dallera}},
  \bibinfo{author}{\bibfnamefont{M.}~\bibnamefont{Grioni}},
  \bibinfo{author}{\bibfnamefont{B.}~\bibnamefont{Delley}},
  \bibinfo{author}{\bibfnamefont{F.}~\bibnamefont{Vernay}},
  \bibinfo{author}{\bibfnamefont{J.}~\bibnamefont{Mesot}},
  \bibinfo{author}{\bibfnamefont{M.}~\bibnamefont{Sikora}},
  \bibinfo{author}{\bibfnamefont{J.~A.} \bibnamefont{Alonso}},
  \bibnamefont{and} \bibinfo{author}{\bibfnamefont{M.~J.}
  \bibnamefont{Martinez-Lope}}, \bibinfo{journal}{Physical Review B}
  \textbf{\bibinfo{volume}{80}}, \bibinfo{pages}{245105}
  (\bibinfo{year}{2009}).

\bibitem[{\citenamefont{Jaramillo et~al.}(2014)\citenamefont{Jaramillo, Ha,
  Silevitch, and Ramanathan}}]{Jaramillo:2014db}
\bibinfo{author}{\bibfnamefont{R.}~\bibnamefont{Jaramillo}},
  \bibinfo{author}{\bibfnamefont{S.~D.} \bibnamefont{Ha}},
  \bibinfo{author}{\bibfnamefont{D.~M.} \bibnamefont{Silevitch}},
  \bibnamefont{and}
  \bibinfo{author}{\bibfnamefont{S.}~\bibnamefont{Ramanathan}},
  \bibinfo{journal}{Nature Physics} \textbf{\bibinfo{volume}{10}},
  \bibinfo{pages}{304} (\bibinfo{year}{2014}).

\end{thebibliography}






% \end{thebibliography}

\end{document}